\documentclass[twocolumn,showpacs]{revtex4}
\usepackage{times,xspace}
\usepackage{amsbsy,amssymb,amsmath,bm}
\usepackage{graphicx,color,epsfig,rotate}
\usepackage{fancyheadings}

\def\bbbc{{\mathchoice {\setbox0=\hbox{$\displaystyle\rm C$}\hbox{\hbox
to0pt{\kern0.4\wd0\vrule height0.9\ht0\hss}\box0}}
{\setbox0=\hbox{$\textstyle\rm C$}\hbox{\hbox
to0pt{\kern0.4\wd0\vrule height0.9\ht0\hss}\box0}}
{\setbox0=\hbox{$\scriptstyle\rm C$}\hbox{\hbox
to0pt{\kern0.4\wd0\vrule height0.9\ht0\hss}\box0}}
{\setbox0=\hbox{$\scriptscriptstyle\rm C$}\hbox{\hbox
to0pt{\kern0.4\wd0\vrule height0.9\ht0\hss}\box0}}}}

\newcommand{\ignore}[1]{}
\newcommand{\mComment}[1]{}
\newcommand{\gComment}[1]{}
\newcommand{\jComment}[1]{}
\newcommand{\rComment}[1]{}
\newcommand{\lComment}[1]{}

\renewcommand{\mComment}[1]{\textcolor{blue}{Manny: #1}}
\renewcommand{\gComment}[1]{\textcolor{red}{Gerardo: #1}}
\renewcommand{\jComment}[1]{\textcolor{green}{Jim: #1}}
\renewcommand{\rComment}[1]{\textcolor{magenta}{Ray: #1}}
\renewcommand{\lComment}[1]{\textcolor{purple}{Rolando: #1}}

\begin{document}

\title{Exact ground states of a frustrated 2D magnet: deconfined 
fractional excitations at a first order quantum phase transition}
\author{C. D. Batista and S. A. Trugman}
\address{Theoretical Division,
Los Alamos National Laboratory, Los Alamos, NM 87545}

\date{Received \today }

\begin{abstract}
We introduce a frustrated spin $1/2$ Hamiltonian which is an extension of the 
two dimensional  $J_1 - J_2$ Heisenberg model. The ground states of this model
are exactly obtained at a first order quantum phase transition between 
two regions with different valence bond solid order parameters. At this point,
the low energy excitations are deconfined spinons and  spin-charge separation 
occurs under doping in the limit of low concentration of holes. In addition,
this point is characterized by the proliferation of topological defects that 
signal the emergence of $Z_2$ gauge symmetry. 
 
\end{abstract}

\pacs{71.27.+a, 71.28.+d, 77.80.-e}

\maketitle

Frustrated magnets are the focus of considerable attention 
because exotic quantum effects are expected to emerge from the 
competition between two or more opposite tendencies. While several
models in this category are solvable in one dimension, the list is much smaller
for higher dimensions. One of the most studied frustrated magnets is the 
spin $1/2$ Heisenberg model with first and second nearest neighbor interactions $J_1$ and 
$J_2$. In one dimension, this model exhibits a quantum transition as 
a function of $J_2/J_1$ from a critical state with quasi-long range antiferromagnetic 
(AF) order to a dimerized phase. Moreover, the exact dimerized ground state 
has been obtained for the point $J_2/J_1=0.5$ by Majumdar and Ghosh \cite{Majumdar}.
In contrast, two dimensional (2D) frustrated magnets like the $J_1-J_2$ Heisenberg model on 
a square lattice still hold many secrets. Different approaches 
predict a transition between a N\'eel ordered state and 
a gapped (non-magnetic) quantum phase for the region $0.4 \lesssim J_2/J_1 \lesssim 0.6$.
However, the nature of this phase is still debated. More precisely, the question is whether it 
is a uniform spin liquid \cite{Fazekas,Capriotti} or a spatially ordered 
valence bond crystal \cite{Read,Kotov,Sushkov,Croo}.
      
The interest in frustration and magnetism is motivated not only by
this widely debated question. There are reasons to believe that frustrated magnets 
may exhibit fractionalized excitations similar to those which appear in the fractional 
quantum Hall effect. The interest in phases with fractionalized excitations is generated by the 
increasing number of experimental results showing new physical behaviors in 
strongly correlated systems. For instance, the normal state of the 
high temperature superconductors does not exhibit electron-like quasiparticles in 
its spectrum. Recently, different scenarios were proposed for the realization of 
points with deconfined fractional excitations. Senthil {\it et al} \cite{Senthil} proposed that 
{\it deconfined critical points} may describe quantum phase transitions between N\'eel ordered states
and valence bond solids (VBS) in frustrated magnets. In a second scenario based on the study of 
models for quantum dimers, the deconfinement point separates two VBS and the spectrum of fractional
excitations consists of spinless particles (``photons'') with a quadratic dispersion 
\cite{Moessner}. Very recently, Tsvelik \cite{Tsvelik} argued that none of the previous 
scenarios is realized for a family of frustrated spin Hamiltonians  called confederate 
flag models \cite{Nersesyan}. As we will see below, his alternative scenario for 
the deconfinement point at the transition between two VBS has many analogies with the case
that we study in the present paper.

In addition, during the last decade attention has focused on the 
study of inhomogeneous structures that are proposed to emerge from competing 
interactions. These textures are generated by the proliferation 
and eventual ordering of one dimensional (1D) topological defects.
A prominent example is provided by the stripe phase proposed to exist
in the high temperature superconductors. In this case, each stripe is
an anti-phase boundary for the antiferromagnetic order parameter.
In general, it is difficult to prove the existence and understand 
the origin of these phases due to the complexity of the underlying 
frustrated model.


In this paper, we will consider the $J_1-J_2$ Heisenberg Hamiltonian on a square lattice
with an additional term that makes the model quasi-exactly solvable for the fully 
frustrated point $J_2/J_1=0.5$. At this point, some of the exact ground states
are valence bond crystals with deconfined fractional $S=1/2$ excitations (spinons).  
In addition, spin-charge separation occurs if the system is doped with a low concentration of holes.
We also show that this particular point has an emergent $Z_2$ gauge symmetry \cite{Batista} and 
can be associated with a {\it first order quantum phase transition} between two different 
valence bond orderings. The physical manifestation of the emergent gauge symmetry is 
a divergent susceptibility for the creation of 1D topological defects that 
can be identified with twin-boundaries of an underlying orientational ordering. 
We will see that the common origin of these exotic behaviors is a dynamical 
decoupling of the 2D magnet into 1D structures.




We will start by considering the following $S=1/2$ Hamiltonian on a square lattice: 
\begin{eqnarray}
H &=& J_1 \sum_{\langle {\bf i}, {\bf j}  \rangle} {\bf S}_{\bf i} \cdot {\bf S}_{\bf j}
+J_2 \sum_{\langle \langle {\bf i}, {\bf j} \rangle \rangle} {\bf S}_{\bf i} \cdot {\bf S}_{\bf j}
\nonumber \\
&+&  K \sum_{\boldsymbol {\alpha}} (P^{\boldsymbol {\alpha}}_{ij} P^{\boldsymbol {\alpha}}_{kl}
+P^{\boldsymbol {\alpha}}_{jk} P^{\boldsymbol {\alpha}}_{il} 
+P^{\boldsymbol {\alpha}}_{ik} P^{\boldsymbol {\alpha}}_{jl}),
\label{Hamil}
\end{eqnarray}
where $\langle {\bf i}, {\bf j}  \rangle$ and 
$\langle \langle {\bf i}, {\bf j} \rangle \rangle$ denote nearest neighbors
and second nearest neighbors respectively. The index ${\alpha}$ denotes the 
sites of the dual lattice (plaquettes) and $ijkl$ are the four spins of the 
corresponding plaquette in cyclic order. Note that the plaquette interaction of $H$ is 
similar to the one introduced by a four cyclic exchange
(the only difference is in the sign of the last term) \cite{Muller}. 
In particular, we will consider here the fully frustrated point $J_2=J_1/2$ and $K=J_1/8$. 
For this set of parameters and up to an irrelevant constant
the Hamiltonian $H$ can be rewritten as:  
\begin{eqnarray}
H_p =  \frac{3 J_1}{2} \sum_{\boldsymbol {\alpha}} {\cal P}^{\boldsymbol {\alpha}},
\label{Hamil2}
\end{eqnarray}
The operator ${\cal P}^{\boldsymbol {\alpha}}$ projects 
the spin state of the plaquette $\boldsymbol {\alpha}$ onto the subspace with 
total spin $S^{\boldsymbol {\alpha}}_T=2$.

{\it Exact ground states of $H_p$.} It is clear that any state having at least one
singlet dimer per plaquette is a ground state of $H_p$. This is because 
$H_p$ is positive semi-definite and the condition of at least one singlet in the plaquette
$\boldsymbol {\alpha}$ implies that $S^{\boldsymbol {\alpha}}_T \leq 1$ 
(there is no $S^{\boldsymbol {\alpha}}_T=2$ component). The same procedure 
gives rise to the Majumdar-Ghosh \cite{Majumdar} and the Affleck-Kennedy-Lieb-Tasaki (AKLT)
\cite{Affleck} exact ground states for $S=1/2$ and $S=1$ chains respectively, and to 
more general ideas for constructing solvable 2D models \cite{Shastry,Sutherland}.

\begin{figure}[htb]
\vspace*{-0.3cm}
\includegraphics[angle=0,height=9cm,width=8.5cm,scale=1.0]{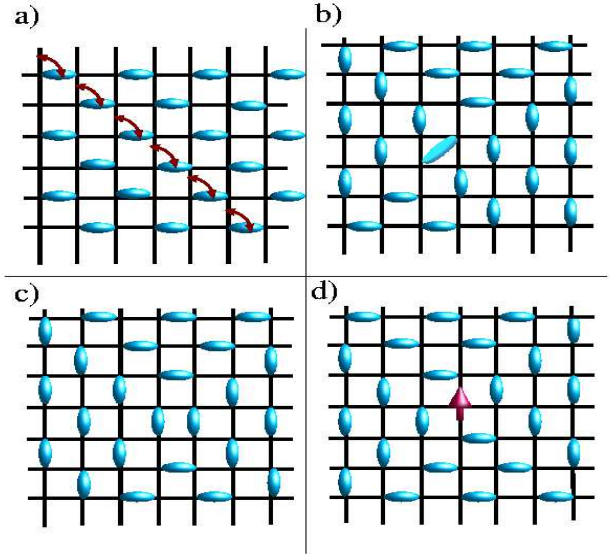}
\vspace*{-1cm}
\caption{Different ground states of $H_p$. The ellipsoids represent singlet dimers.}
\label{fig1}
\end{figure}

We found two different families of states that fulfill the condition
of having at least one singlet dimer per plaquette. The first family is generated by
the state which is illustrated in Fig.\ref{fig1}a. These states are simply
products of local singlet dimers which are represented by ellipsoids. 
In other words, the singlet dimers are completely localized and there is 
emergent $U(1)$ gauge symmetry associated with this localization \cite{Batista}.
Any array of dimers along a given diagonal direction, $(1,1)$ or $(1,{\bar 1})$, can be rotated
by $\pi/2$ as indicated by the arrows of Fig.\ref{fig1}a. It is important to note
that successive rotations have to be done along the the same diagonal direction.
The degeneracy of this family is $2^{N_d+1}$ where 
$N_d \propto \sqrt{N_s}$ is the number of diagonal chains and $N_s$ is the total number of sites.
This degeneracy can be associated with a $Z_2$ gauge {\it emergent symmetry} that changes 
the dimerized order parameter of each individual diagonal zig-zag chain. By $Z_2$ gauge symmetry
we mean a local symmetry transformation that acts on each individual zig-zag chain
mapping one of the two possible dimerized states into the other one. 
In other words, this family of ground states is formed by configurations with parallel diagonal 
arrays of vertical or horizontal dimers.

There is an alternative way of viewing the local $Z_2$ emergent symmetry.
We can imagine that our system has an underlying orientational ordering given by
the staggered dimer phase illustrated in Fig.\ref{fig1}a. The two possible 
orientations are horizontal and vertical. The energy cost for creating a 
twin-boundary or inter-phase between the two different orientations is equal to zero.
Consequently, the system has a divergent susceptibility for the 
creation of 1D topological defects, meaning that a weak 
coupling with another field can easily stabilize a particular array of twin boundaries.
If the system is still invariant under translations, the resulting array
of topological defects will be periodic. The ordering of 1D topological defects
has been proposed in the past as a possible outcome of competing interactions 
in the high temperature superconductors \cite{Zaanen}.

The second family (Figs.\ref{fig1}b, c and d) appears when we force the states 
to have interfaces between vertical and horizontal configurations along 
the two diagonal directions $(1,1)$ and $(1,{\bar 1})$. Under this condition
the state is forced to create a defect at the intersection between the two 
diagonal interfaces. In particular, it is possible to have an $S=1/2$ defect
or localized {\it spinon} as is illustrated in Fig.\ref{fig1}d. It is easy 
to prove that there is no ground state with more than one localized spinon. Since
the position of the defect is arbitrary, the degeneracy of 
each of these configurations is proportional to $N_s$. 

\begin{figure}[htb]
\vspace*{-0.3cm}
\includegraphics[angle=0,height=11cm,width=8.5cm,scale=1.0]{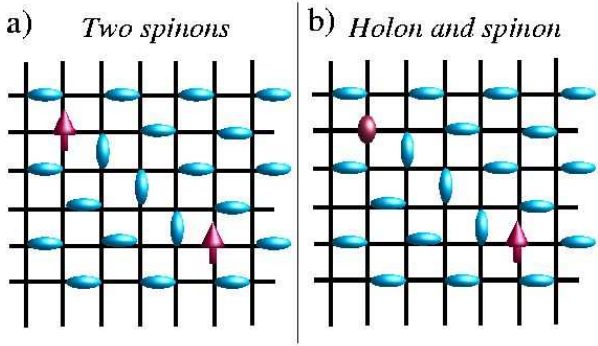}
\vspace*{-5.0cm}
\caption{
\label{fig2}
Deconfined fractional excitations. a) Two deconfined spinons.
b) Spin-charge separation when a low concentration of holes is 
introduced in the system.}
\end{figure}

{\it Deconfined fractional excitations.} What are 
the low energy excitations of the ground states of Fig.\ref{fig1}?
Let us first consider the family of solutions illustrated in  
Fig.\ref{fig1}a. As we can see in Fig.\ref{fig2}a, if we excite one
singlet dimer into a triplet state, the two parallel $S=1/2$ 
spins can propagate along the diagonals without an 
energy cost proportional to the separation between them. Consequently,
the two spinon excitations are {\it deconfined}. Note that when the 
spinons propagate along one diagonal zig-zag chain they do not ``see'' 
the other chains because the two possible dimerized configurations
(horizontal and vertical) have exactly the same energy. In other words, 
the effective dimensionality  of the low energy spectrum is dynamically reduced from D=2 to D=1. 
The emergent $Z_2$ gauge symmetry that we mentioned above is the mathematical
manifestation of this dynamical decoupling. The $Z_2$ dimerized order parameter 
of each diagonal zig-zag chain is decoupled from the corresponding order parameter of the 
other chains. The most notorious physical 
consequence is the emergence of fractional excitations which are 
characteristic of 1D systems.

Note that the previous analysis is only valid when the spinons are
moving along the two diagonal directions. If, for instance, 
they propagate horizontally, they will ``feel'' the confining interaction 
characteristic of 2D systems because the horizontal chains are not dynamically decoupled. Consequently,
although our system is 2D, its low energy spinon excitations
are free to move only along 1D paths. The same type of deconfined spinon excitations
are obtained for the second family of ground states.

A similar situation occurs when one hole is introduced (Fig.\ref{fig2}b). 
The charge and the spin of the added hole get deconfined because
there is no energy cost for the string generated in between the two excitations.
In other words, we can expect a non-Fermi liquid behavior when a magnetic system 
described by $H_p$ is doped away from half-filling. All of these ``exotic'' behaviors 
are just a consequence of a dynamical decoupling which is signaled by an emergent
$Z_2$ gauge symmetry and that only occurs at the point under consideration: 
$J_2=J_1/2$ and $K=J_1/8$. To understand what can be the physical role of this point 
we need to move away from it. 

\begin{figure}[htb]
\vspace*{-0.3cm}
\includegraphics[angle=0,height=10cm,width=8.5cm,scale=1.0]{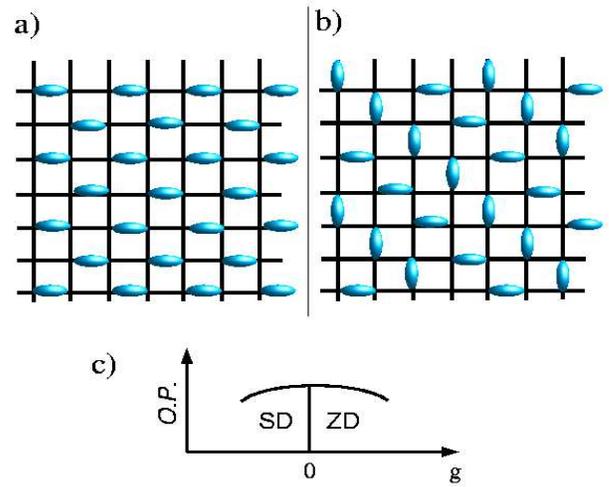}
\vspace*{-3.0cm}
\caption{
\label{fig3}
a) Staggered dimer phase (SD). b) Zig-zag dimer phase (ZD). c) The two order parameters
change discontinously at $g=0$ indicating a first order 
quantum phase transition between the phases illustrated in a) and b).}
\end{figure}

{\it First order quantum phase transition.} The special point 
described by $H_p$ can be easily converted into a {\it first order}
quantum phase transition point. Note that the two configurations of 
Fig.\ref{fig3} are the periodic ground states with the shortest periods 
within the set shown in Fig.\ref{fig1}. 
Therefore, they are the leading candidates to remain as ground states when we depart from 
the $H_p$ point. For instance, we can add a term to the Hamiltonian which favors the 
{\it staggered} dimer ordering shown in Fig.\ref{fig3}a when the coupling constant 
$g$ is negative or the {\it zig-zag} dimer configuration of Fig.\ref{fig3}b when $g$ 
is positive. There are different realizations of such a term
and we will not focus in any particular one. The two different 
dimer phases shown in Fig.\ref{fig3} break simultaneously the 
translation and the rotation symmetry of the square lattice. 
The first dimer configuration (Fig.\ref{fig3}a) is four-fold degenerate while the 
second one (Fig.\ref{fig3}b) has an eight-fold degeneracy.
The symmetry order parameters for each of these phases are:
\begin{equation}
{\cal D}^{st}_{\eta}= \frac{1}{N_s} \sum_{\bf j} 
({\bf S}_{\bf j} \cdot {\bf S}_{{\bf j}+{\hat {\eta}}})
e^{i{\bf Q}\cdot{\bf r_j}},
\end{equation}
for the staggered dimer ordering of Fig.\ref{fig3}a and 
\begin{eqnarray}
{\cal D}^{zz}_{x}= \frac{1}{N_s} \sum_{\bf j} 
({\bf S}_{\bf j} \cdot {\bf S}_{{\bf j}+{\hat {x}}}
+ {\bf S}_{\bf j+2{\hat { x}}} \cdot 
{\bf S}_{{\bf j}+2{\hat x}+{\hat y}})
e^{\frac{i}{2}{\bf Q}\cdot{\bf r_j}},
\nonumber \\
{\cal D}^{zz}_{y}= \frac{1}{N_s} \sum_{\bf j} 
({\bf S}_{\bf j} \cdot {\bf S}_{{\bf j}+{\hat {y}}}
+ {\bf S}_{\bf j+2{\hat {y}}} \cdot 
{\bf S}_{{\bf j}+2{\hat x}+{\bar y}})
e^{\frac{i}{2}{\bf {\bar Q}}\cdot{\bf r_j}},
\end{eqnarray}
where ${\bf Q}=(\pi,\pi)$, ${\bf {\bar Q}}=(\pi,-\pi)$, and
${\eta}=\{{\hat x},{\hat y} \}$. Note that 
$({\cal D}^{st}_{x},{\cal D}^{st}_{y})=(\pm 1,0)$ or $(0,\pm 1)$ for the 
four equivalent ground states of the staggered dimer phase. For the 
zig-zag phase, the non-zero component of $({\cal D}^{zz}_{x},{\cal D}^{zz}_{y})$
takes the four possible values $\{ 1, i, -1, -i \}$ that are necessary to
identify the eight equivalent configurations. The remaining symmetry group
of the zig-zag phase, ${{\cal G}^{zz}}$, is a subgroup of ${{\cal G}^{st}}$, the 
symmetry group of the staggered dimer phase. 
     The level crossing that occurs at $g=0$ between the staggered and the zig-zag dimer 
states gives rise to a first order quantum phase transition. In other words, at this point
there is a discontinuous change of the order parameters ${\cal D}^{st}$ and 
${\cal D}^{zz}$  (see Fig.\ref{fig3}c). As we mentioned above, this level crossing
is accompanied by the softening of the twin-boundary defects of the staggered dimer 
phase. Hence, we can think of the zig-zag dimer phase as a ``condensation" of 
these twin-boundaries. 


What is the general feature that gives rise to these exotic behaviors?  
To answer this question it is convenient to think of our system as an array of diagonal
zig-zag chains. For each of these chains we can introduce the usual 
$Z_2$ dimer order parameter ${\cal D}_l$, where $l$ is the chain index. 
We can now build different order parameters for the 2D system
by choosing different periodic configurations of $D_l$ with a well defined inter-chain
wave vector $k$. For instance, in the case under consideration the dimer phase 
of  Fig.\ref{fig3}a corresponds to the uniform ($k=0$) configuration ${\cal D}_l=D$, 
i.e., the one-dimensional $Z_2$ order parameter points in the same direction for 
all the different chains. In contrast, the zig-zag dimer 
ordering of Fig.\ref{fig3}b corresponds to ${\cal D}_l=(-1)^l D$, i.e., there
is a staggered ($k=\pi$) inter-chain ordering. Since the different states built in this way
have different wave vectors $k$, the quantum phase transition between two of
them is of first order. At the transition point, both states have the same energy
meaning that there are adjacent chains for which the two relative orientations of 
$D_l$ are degenerate. In other words, the effective coupling between these two 
chains has been reduced to zero and a $Z_2$ gauge symmetry emerges at the transition point.
In general, we can say that these are quantum phase transitions between two 
broken symmetry states whose order parameters  have something in common: 
given a particular decomposition of our 2D system into 1D chains
described by some macroscopic variable, both order parameters 
characterize different inter-chain orderings. 

For real systems, we do not expect the dynamical decoupling into 1D systems
to be perfect. In general, there is always some residual interaction that
makes this coupling weak but non-zero. A similar situation occurs in the
materials that provide an experimental realization of one dimensional systems. 
The structure of these materials contains weakly coupled chains and the coupling becomes 
relevant when temperature is lower than some characteristic value $T_0$ . 
However, the system still behaves as an array of decoupled chains for $T>T_0$.   
We expect the same behavior for 2D systems that get dynamically deconfined into
1D strucutures. 

These ideas can be extended to other lattices. For instance, for a 
honeycomb lattice we can consider a Hamiltonian that is the sum 
of operators ${\cal Q}^{\boldsymbol \alpha}$ that project the 
the spin state of the plaquette ${\boldsymbol \alpha}$ onto the 
subspace with total spin $S_T^{\boldsymbol \alpha}=2,3$. As in the previous 
case, the system gets dynamically deconfined into chains that have two 
possible dimerized states. Consequently there is an 
emergent $Z_2$ gauge symmetry that gives rise to the exotic behaviors 
which are described above.

{\it Conclusions.} In summary, we introduced a simple extension of the 
$J_1-J_2$ Heisenberg model and we obtained different exact ground states 
for the fully frustrated point $J_2/J_1=0.5$, $K=J_1/8$. Both
the ground states and their low energy excitations exhibit 
exotic behaviors like the softening of 1D toplogical defects and 
the emergence of deconfined fractional excitations. When some of the spins are 
replaced by holes, the phenomenon of spin-charge separation occurs in the limit of low concentration 
of holes. In addition, we showed that the point under consideration can be 
easily identified with a first order quantum phase transition.

The emergence of deconfined fractional excitations was recently proposed to occur
at a  critical point by different authors \cite{Senthil,Moessner,Tsvelik}. In particular,
our model is an isotropic version of the anisotropic Confederate Flag model studied by
Tsvelik \cite{Tsvelik}. By analyzing the four chain model, he finds that there are two VBS 
separated by an approximately (1+1) Lorentz invariant quantum critical point.
However, the exponent that he obtains for the average dimerization is quite small,
indicating that in the limit of infinite number of chains the transition could become 
first order as in the isotropic model discussed in this paper.
Our results then show that this phenomenon is
not restricted to quantum critical points. First order quantum phase transition points 
can also produce strong deviations from the normal properties of 2D systems. 
In our case, the common origin of these deviations is a dynamical decoupling of the 
2D magnet into 1D systems. Such a dynamical decoupling was proposed for 2D 
strongly correlated models in the context of the high temperature superconductors
\cite{Anderson}. As is well known, this effective reduction in the dimensionality 
of the system gives rise to completely different properties: the usual Fermi liquid is
replaced by a Luttinger liquid and the magnetic spectrum is dominated by $S=1/2$ spinon
excitations.

This work was sponsored by the US DOE under contract
W-7405-ENG-36, PICT 03-06343 of ANPCyT.

\end{document}